\documentstyle[12pt]{article}
\topmargin - 1.25cm
\begin{document}
\newcommand{\bea}{\begin{eqnarray}}
\newcommand{\eea}{\end{eqnarray}}
\newcommand{\be}{\begin{equation}}
\newcommand{\ee}{\end{equation}}
\newcommand{\non}{\nonumber}
\global\parskip 6pt
\begin{titlepage}
\vspace*{1cm}
\begin{center}
{\Large\bf A Closed Contour of Integration in Regge Calculus}
\vskip .50in
Danny Birmingham \footnote{Email: Dannyb@ollamh.ucd.ie}
\vskip .10in
{\em University College Dublin, Department of Mathematical Physics,\\
Belfield, Dublin 4, Ireland} \\
\end{center}
\vskip .10in
\begin{abstract}
The analytic structure of the Regge action on a cone in $d$ dimensions
over a boundary of arbitrary topology is determined in simplicial
minisuperspace. The minisuperspace is defined by the assignment of a
single internal edge length to all $1$-simplices emanating from the
cone vertex, and a single boundary edge length to all $1$-simplices
lying on the boundary. The Regge action is analyzed in the space of
complex edge lengths, and it is shown that there are three finite
branch points in this complex plane. A closed contour of integration
encircling the branch points is shown to yield a convergent real wave
function. This closed contour can be deformed to a steepest descent
contour for all sizes of the bounding universe. In general, the contour
yields an oscillating wave function for universes of size greater than
a critical value which depends on the topology of the bounding universe.
For values less than the critical value the wave function exhibits
exponential behaviour. It is shown that the critical value is positive
for spherical topology in arbitrary dimensions. In three dimensions we
compute the critical value for a boundary universe of arbitrary genus,
while in four and five dimensions we study examples of product manifolds
and connected sums.
\end{abstract}
\vskip 1cm
\begin{center}
October 1997
\end{center}
\end{titlepage}

\section{Introduction}
Regge calculus provides a framework in which one can study a simplicial
approach to the quantization of the gravitational
field \cite{Regge}-\cite{Hamber}. The basic
idea is to model the spacetime of interest by a simplicial complex,
in which the edge length assignments become the dynamical variables.
A central question in any approach to such a problem is how to define
the associated functional integral. In particular,
one must deal with the well-known convergence
problems of the corresponding continuum Euclidean functional
integral \cite{GHP}.

In \cite{Hartle2}, the functional integral for a simplicial complex
with a single boundary component of $S^{3}$ topology was studied.
It was shown that by taking the dynamical variables to be given
by the space of complex valued edge lengths, one could
study the convergence and physical properties
of the associated Hartle-Hawking wave function \cite{HarHaw}
in explicit detail.
One simplifying feature introduced was to restrict attention to
a simplicial minisuperspace. This minisuperspace consisted
of the assignment of a single internal edge length variable,
and a single boundary edge length variable to the simplicial
complex.
The functional integral then reduced to an integral over
a single complex variable. In particular, it was shown that
the Regge action had three square root branch points,
and that a closed contour of integration encircling these points
led to a convergent result. More particularly, this contour had
the appealing feature that the form of the resulting wave function
satisfied certain physical requirements. The general criteria
for defining the wave function of the universe were examined
in \cite{HallHar}.
For values of the boundary edge length greater than a critical
value, the wave function behaved in an oscillatory
manner (corresponding to a classically allowed regime),
and was dominated semiclassically by real
simplicial geometries of Lorentzian signature.
For values of the boundary edge length less
than the critical value, the wave function behaved in
an exponential way (corresponding to a classically forbidden
regime), with a  semiclassical
domination by real geometries of
Euclidean signature.

In \cite{Bir1,Bir2},
this result was extended to a class of non-trivial topologies,
given by the lens spaces $L(k,1)$. The essential ingredient in
this generalization was the observation that
the features in \cite{Hartle2} which allowed the analysis to proceed
were the particular simplicial minisuperspace considered, and
the fact that the spacetime simplicial complex was chosen to be a cone
over $S^{3}$.

In this paper, we consider Regge calculus on a cone in arbitrary
dimensions over a boundary of arbitrary topology.
The basic object of interest is the corresponding functional
integral which yields the Hartle-Hawking
wave function for a universe with the
boundary topology. For a simplicial complex with the structure
of a cone, there is a single internal vertex, and we study
the model for a simplicial minisuperspace in which all edge
lengths which
emanate from this cone vertex take equal value. In addition, we
take all boundary edge lengths to be equal. As in the cases discussed
above, we study the functional integral in the space of complex
valued edge lengths. The basic result is that due
to the cone structure and simplicial minisuperspace,
the analytic structure of the Regge action
parallels that studied in the four-dimensional case.
In particular, there are three square root branch points,
and consequently we can appeal to the general argument presented in
\cite{Hartle2}.
We remark here that the simplicial cone structure is not
a simplicial manifold, as defined in section three. Nevertheless,
the formalism of Regge calculus is defined in this more general setting.
Our aim is to obtain the Hartle-Hawking wave function of
the associated boundary universe, and study its physical properties
for a wide variety of topologies.
In particular, we evaluate the critical edge length value
which allows us to determine the classically forbidden
and allowed regimes.

The outline of this work is as follows. In the following section,
we determine the analytic structure of the Regge action
for a simplicial complex which is a cone over a boundary
of arbitrary topology in $d$ dimensions.
We find that the action has three square root branch points,
and the classical extrema of the action are obtained.
It is shown that for boundary edge lengths less than a critical
value, there are real simplicial geometries of Euclidean signature,
and for boundary edge lengths greater than the critical value,
there are classical extrema corresponding to real geometries
of Lorentzian signature.
The analytic structure of the action is shown to parallel
that found in \cite{Hartle2},
and thus a closed contour of integration encircling
all three branch points yields a convergent wave function,
with the desired physical properties.
Following this general analysis, we then study particular examples
in specific dimensions. We also determine the form of the wave
function for a universe with spherical topology in arbitrary dimensions.

\section{The Regge Action on a Cone}
Given a simplicial complex $M_{d}$ with boundary $\partial M_{d} = M_{d-1}$,
the wave function is defined by
\be
\Psi(s_{b}) = \int_{C} \; d\mu(s_{i})\;\exp[-S(s_{b},s_{i})].
\label{r1}
\ee
Here, the variables $s_{b}$ specify the edge lengths of the boundary
and the integration is over the internal dynamical variables $s_{i}$.
The form of the measure $\mu$, the contour of integration $C$, along
with the action $S$, are required to complete the
specification of the model.

The Euclidean Einstein action with cosmological term for
a manifold with boundary is given by
\be
S = - \frac{1}{16 \pi G} \int_{M} d^{d}x \;\sqrt{g} R +
\frac{2\Lambda}{16\pi G}\int_{M} d^{d} x \;\sqrt{g}
- \frac{2}{16 \pi G} \int_{\partial M} d^{(d-1)} x\; \sqrt{h} K,
\label{r2}
\ee
where $R$ is the scalar curvature of the metric $g$, 
$\Lambda$ is the cosmological constant, 
and $K$  is the extrinsic curvature scalar of the induced metric $h$ on
the boundary. 
The simplicial analogue of this action is the 
corresponding Regge action \cite{Regge,HarSor},
which is given by
\bea
S &=& -\frac{2}{l^{d-2}}\sum_{\sigma_{d-2} \subset
int(M_{d})} V_{d-2}
(\sigma_{d-2})
\theta(\sigma_{d-2}) + \frac{2 \Lambda}{l^{d-2}}\sum_{\sigma_{d}
\subset int(M_{d})}V_{d}(\sigma_{d}) \non\\
&-& \frac{2}{l^{d-2}}\sum_{\sigma_{d-2} \subset \partial M_{d}} V_{d-2}
(\sigma_{d-2})\psi(\sigma_{d-2}),
\label{r3}
\eea
where the Planck length, in units where $\hbar = c = 1$, 
is $l = (16 \pi G)^{1/(d-2)}$. The various terms in (\ref{r3}) are described
as follows.
The Einstein term is represented by a summation over internal
$(d-2)$-simplices
$\sigma_{d-2}\subset int(M_{d})$ (also known as hinges). 
An internal hinge is any $(d-2)$-simplex
of the complex which contains at least one internal vertex, 
and the notation
$int(M_{d})$ is used to denote this set. The form of
the Einstein action involves the volume of the hinge $V_{d-2}(\sigma_{d-2})$
and the associated
deficit angle $\theta(\sigma_{d-2})$. 
Similarly, the boundary term is given in terms
of the boundary $(d-2)$-simplices and their associated deficit angles
denoted by $\psi(\sigma_{d-2})$.
The cosmological term is simply represented as a sum over the volumes
$V_{d}(\sigma_{d})$ of the $d$-simplices $\sigma_{d}$ 
of the complex.

The deficit angle of an internal $(d-2)$-simplex $\sigma_{d-2}^{int}$
is given by
\be
\theta(\sigma_{d-2}^{int}) =
2 \pi - \sum_{\sigma_{d} \supset \sigma_{d-2}^{int}}
\theta_{dih}(\sigma_{d-2}^{int},\sigma_{d}) \;\;,
\label{r4}
\ee
where the summation is over
the dihedral angles of all $d$-simplices containing
$\sigma_{d-2}^{int}$.
The deficit angle for the boundary
$(d-2)$-simplex $\sigma_{d-2}^{bdy}$ is
\be
\psi(\sigma_{d-2}^{bdy}) =
\pi - \sum_{\sigma_{d} \supset \sigma_{d-2}^{bdy}}
\theta_{dih}(\sigma_{d-2}^{bdy},\sigma_{d})\;\;.
\label{r5}
\ee

A simple way to obtain a simplicial complex with a boundary
is to take the simplicial complex to be the cone over the boundary
\cite{Munkres}. The simplicial cones considered here belong to the general
class of simplicial conifolds discussed in \cite{SchWitt}.
Given a $(d-1)$-dimensional simplicial complex $M_{d-1}$, the cone over
$M_{d-1}$ is denoted by $M_{d} = c \star M_{d-1}$ and
is described as follows \cite{Munkres}.
One takes the vertex set of $M_{d-1}$ and adds the cone vertex $c$.
The cone vertex is then joined to each of the vertices of $M_{d-1}$.
Each $d$-simplex of the cone is
then of the form $[c,\sigma_{d-1}]$, where $\sigma_{d-1}$ is
a $(d-1)$-simplex of the boundary complex. With this orientation,
the boundary of the cone is $\partial (c\star M_{d-1}) = M_{d-1}$.
In fact, the number of simplices of each dimension is given by
\bea
N_{0}(M_{d}) &=& N_{0}(M_{d-1}) + 1,\nonumber\\
N_{1}(M_{d}) &=& N_{1}(M_{d-1}) + N_{0}(M_{d-1}),\nonumber\\
&\vdots&\nonumber\\
N_{d-2}(M_{d}) &=& N_{d-2}(M_{d-1}) + N_{d-3}(M_{d-1}),\nonumber\\
N_{d-1}(M_{d}) &=& N_{d-1}(M_{d-1}) + N_{d-2}(M_{d-1}),\nonumber\\
N_{d}(M_{d}) &=& N_{d-1}(M_{d-1}).
\label{r6}
\eea
Thus, we see that there is a total of $N_{d-2}(M_{d-1})$ boundary
$(d-2)$-simplices, and a total of $N_{d-3}(M_{d-1})$ internal
$(d-2)$-simplices.

So let us consider a $d$-simplex of this cone
given by $\sigma_{d} = [c,1,2,...,d]$. We let
$e_{1},e_{2},...,e_{d}$ be $1$-forms associated with the $1$-simplices
$[c,1],...,[c,d]$. The volume $d$-form for $\sigma_{d}$ is then
given by $\omega_{d} = e_{1}\wedge ... \wedge e_{d}$. The
volume of $\sigma_{d}$ is
\be
V_{d}(\sigma_{d}) = \frac{1}{d!} Det^{\frac{1}{2}}(M_{d}),
\label{r7}
\ee
where $M_{d} \equiv \omega_{d}\cdot\omega_{d}$ has entries
$m_{ij} = e_{i}\cdot e_{j} = \frac{1}{2}(s_{ci} + s_{cj} - s_{ij})$,
and $s_{ij}$ is the squared edge length assigned to
the $1$-simplex $[i,j]$, see for example \cite{Hartle1}.

In order to compute the internal deficit angles, we need a formula
for the dihedral angle of the internal $(d-2)$-simplex
$\sigma_{d-2}^{int} = [c,1,2,...,d-2]$ which is contained in $\sigma_{d}$.
This $(d-2)$-simplex is shared
by the two $(d-1)$-simplices $\sigma_{d-1} = [c,1,2,...,d-2,d-1]$
and $\sigma_{d-1}^{\prime} = [c,1,2,...,d-2,d]$.
The volume forms of $\sigma_{d-1}$ and $\sigma_{d-1}^{\prime}$ are
\bea
\omega_{d-1} &=& e_{1}\wedge ...\wedge e_{d-2}\wedge e_{d-1},\non\\
\omega_{d-1}^{\prime} &=& e_{1} \wedge ... \wedge e_{d-2} \wedge e_{d}.
\label{r8}
\eea
The dihedral angle is
\be
\cos \theta_{dih}(\sigma_{d-2}^{int},\sigma_{d}) =
\frac{1}{((d-1)!)^{2}}
\frac{Det(M_{d-1})}{V_{d-1}(\sigma_{d-1})
V_{d-1}(\sigma_{d-1}^{\prime})},
\label{r9}
\ee
where $M_{d-1} = \omega_{d-1}\cdot \omega^{\prime}_{d-1}$.

For the boundary dihedral angle, we consider the $(d-2)$-simplex
$\sigma_{d-2}^{bdy} = [1,2,...,d-1]$. We choose an orientation
such that the vertices $1,2,...,d-1$ appear at the
beginning of the two $(d-1)$-simplices as
$\sigma_{d-1} = [1,2,...,d-1,c]$ and $\sigma_{d-1}^{\prime} =
[1,2,...,d-1,d]$.
The volume forms of $\sigma_{d-1}$ and $\sigma_{d-1}^{\prime}$
are
\bea
\omega_{d-1} &=& (e_{2} - e_{1})\wedge (e_{3} - e_{1}) \wedge ...
\wedge (e_{d-1} - e_{1}) \wedge (-e_{1}),\non\\
\omega_{d-1}^{\prime} &=& (e_{2} - e_{1}) \wedge (e_{3} - e_{1})
\wedge ... \wedge (e_{d-1} - e_{1}) \wedge (e_{d} - e_{1}).
\label{r10}
\eea
The dihedral angle is then
\be
\cos \theta_{dih}(\sigma_{d-2}^{bdy}, \sigma_{d}) =
\frac{1}{((d-1)!)^{2}}\frac{Det(M_{d-1})}
{V_{d-1}(\sigma_{d-1})V_{d-1}^{\prime}(\sigma_{d-1})}.
\label{r11}
\ee

In order to obtain an explicit expression for the action of
a $d$-dimensional cone, one needs to compute various volumes and deficit
angles. The important point for our purposes is that we can obtain
closed expressions for these in the simplicial minisuperspace
of interest.
We assign an edge length variable $s_{i}$ to each of the
internal $1$-simplices, and an edge length variable $s_{b}$
to each of the boundary $1$-simplices. It is useful to
define the dimensionless ratio $z = s_{i}/s_{b}$.

Let us begin by considering the volume of an internal  $n$-simplex
$\sigma_{n}^{int} = [c,1,2,...,n]$.
From (\ref{r7}), we see that it takes the form
\be
V_{n}(\sigma_{n}^{int}) = \frac{s_{b}^{n/2}}{n!}
Det^{\frac{1}{2}}(M_{n}),
\label{r12}
\ee
where
\bea
M_{n} =
\left(\begin{array}{cccccc}
z & z- \frac{1}{2}& \cdots &z - \frac{1}{2} \\
z- \frac{1}{2} & z &  \cdots & z - \frac{1}{2}\\
\vdots & \vdots & \ddots &  \vdots \\
z- \frac{1}{2} & z- \frac{1}{2} & \cdots & z
\end{array} \right).
\label{r13}
\eea
The determinant can be evaluated explicitly by row reducing
$M_{n}$ to echelon form.
Thus, from (\ref{r12}), the
volume of an internal $n$-simplex is given by
\be
V_{n}(\sigma_{n}^{int}) = \frac{s_{b}^{n/2}}{n!}\sqrt{\frac{n}{2^{n-1}}}
\sqrt{z - \frac{n-1}{2n}}.
\label{r14}
\ee
In order to compute the volume of a boundary $n$-simplex
$\sigma_{n}^{bdy} = [1,...,n]$, we can proceed along similar lines.
However, one notices that the boundary volumes can be obtained
from (\ref{r14}) by setting $z=1$. Thus, the volume of a boundary
$n$-simplex is
\be
V_{n}(\sigma_{n}^{bdy}) = \frac{s_{b}^{n/2}}{n!}
\sqrt{\frac{n+1}{2^{n}}}.
\label{r15}
\ee

The dihedral angles are computed as follows. Consider first
the case of an internal $(d-2)$-simplex $\sigma_{d-2}^{int} =
[c,1,...,d-2]$.
From (\ref{r9}), we see that it is given by
\be
\cos \theta_{dih}(\sigma_{d-2}^{int},\sigma_{d}) = \frac{s_{b}^{d-1}}
{((d-1)!)^{2}}
\frac{Det(M_{d-1})}{V_{d-1}(\sigma_{d-1})V_{d-1}(\sigma_{d-1}^{\prime})},
\label{r16}
\ee
where
\bea
M_{d-1} =
\left(\begin{array}{ccccccc}
z & z - \frac{1}{2}&\cdots & z - \frac{1}{2} \\
z- \frac{1}{2} & z & \cdots & z - \frac{1}{2}\\
\vdots & \vdots & \ddots&\vdots\\
z-\frac{1}{2}& z- \frac{1}{2}&\cdots & z-\frac{1}{2}
\end{array} \right).
\label{r17}
\eea
In this case, one of the volume factors is of internal type
and one is of boundary type.
Thus, the dihedral angle of internal type is given by
\be
\cos \theta_{dih}(\sigma_{d-2}^{int}, \sigma_{d}) =
\frac{z - \frac{1}{2}}
{(d-1)\left(z - \frac{d-2}{2(d-1)}\right)}.
\label{r18}
\ee
Consider now a boundary $(d-2)$-simplex
$\sigma_{d-2}^{bdy} = [1,2,...,d-1]$.
This is shared by the two $(d-1)$-simplices contained
in $\sigma_{d} = [c,1,2,...,d]$, namely
$\sigma_{d-1} = [c,1,2,...,d-1]$ and
$\sigma_{d-1}^{\prime} = [1,2,...,d-1,d]$ which
are of internal and boundary type, respectively.
The boundary dihedral angle is then computed to be
\be
\cos\theta_{dih}(\sigma_{d-2}^{bdy},\sigma_{d}) = \frac{1}
{\sqrt{2d(d-1)\left(z - \frac{d-2}{2(d-1)}\right)}}.
\label{r19}
\ee

With these results at hand, it is straightforward to write down
the complete analytic form of the action. It is convenient to
write it in the form
\be
S(z,R) = \frac{1}{H^{d-2}}\left[-R^{\frac{d}{2}-1} F(z) +
R^{\frac{d}{2}} G(z)\right],
\label{r20}
\ee
where
\bea
F(z) &=& \frac{d N_{d-1}(M_{d-1})}{(d-2)!}\sqrt{\frac{d-1}{2^{d-2}}}
\left[\pi - 2\arccos\left(\frac{1}{\sqrt{2d(d-1)(z - z_{2})}}
\right)\right]\non\\
&+& \left. \frac{2N_{d-3}(M_{d-1})}{(d-2)!} \sqrt{\frac{d-2}{2^{d-3}}}
\sqrt{z - z_{1}}\right[2 \pi \non\\
&-&\left. \frac{d(d-1)N_{d-1}(M_{d-1})}{2N_{d-3}(M_{d-1})}
\arccos\left(\frac{z - \frac{1}{2}}{(d-1)(z-z_{2})}\right) \right],\non\\
G(z) &=& \frac{6N_{d-1}(M_{d-1})}{d!}
\sqrt{\frac{d}{2^{d-1}}}\sqrt{z - z_{3}}.
\label{r21}
\eea
In the above, we have introduced the dimensionless
variable $R = H^{2} s_{b}/l^{2}$, where $H^{2} = l^{2}\Lambda/3$.
It is also convenient to define
\be
z_{1} = \frac{d-3}{2(d-2)},\;\; z_{2}=\frac{d-2}{2(d-1)},\;\;
z_{3} = \frac{d-1}{2d}.
\label{r22}
\ee
We thus see that the action is specified
completely in terms
of the number $N_{d-1}(M_{d-1})$ of $(d-1)$-simplices,
and the number $N_{d-3}(M_{d-1})$ of $(d-3)$-simplices of the boundary.
It is in this sense that the form of the action depends only on simple
data associated with the boundary.

It is important to make the following
observation regarding the factor of $2$ appearing in the 
formula for the deficit angle of a boundary $(d-2)$-simplex (the first term
in $F(z)$). 
The bounding space $M_{d-1}$ is represented by a closed simplicial
complex, and closure of the complex means that each $(d-2)$-simplex is 
contained in  precisely two $(d-1)$-simplices.  
Thus, when we elevate this boundary
complex to its associated cone, we immediately know that the
number of 
$d$-simplices containing each boundary $(d-2)$-simplex is again precisely 
$2$, each being of the form $[c,\sigma_{(d-1)}]$, with $\sigma_{(d-1)}$
belonging to the boundary.  This
fact is crucially relevant when we search for the extrema of
the action.

In computing the Einstein term, we note that each internal
$(d-2)$-simplex is of the form $\sigma^{int}_{d-2} = [c,\sigma_{d-3}^{bdy}]$.
The number of $d$-simplices containing $\sigma^{int}_{d-2}$ is then
equal to the number
of boundary $(d-1)$-simplices containing $\sigma^{bdy}_{d-3}$.
This number depends on the individual $(d-3)$-simplex.
However, because of the structure of the simplicial minisuperspace,
we only need the total number of boundary $(d-3)$-simplices contained
in all the boundary $(d-1)$-simplices.  Bearing in mind that
each boundary $(d-1)$-simplex contains a total
$d(d-1)/2$ of $(d-3)$-simplices, we obtain the Einstein term as given above.

The analytic structure of the action is immediately identified.
We see that there are three finite square root branch points
at $z_{1},z_{2},z_{3}$. It is also clear that it is respectively
the vanishing
of the internal $(d-2)$-volume, internal $(d-1)$-volume,
and internal $d$-volume, which is responsible for the presence of these
branch points. One should also note that there is a logarithmic branch
point at $z_{2}$.

In order to be able to interpret the nature of the wave function,
it is necessary to know the signature of the metric tensor for
each of the $d$-simplices. This is obtained by
finding the eigenvalues of the metric tensor, $M_{d}$ of eqn.
(\ref{r7}), of each $d$-simplex.
Once again, this calculation can be performed by row reduction,
yielding the result that there is one eigenvalue $\lambda = \frac{1}{2}$
with a multiplicity of $(d-1)$, and an eigenvalue
$\lambda = d\left(z - z_{3}\right)$.
Thus, we see that for real $z > z_{3}$, we have a regime of real
Euclidean signature metrics, while for real $z < z_{3}$
we have a regime of real Lorentzian signature metrics.

Given the presence of the branch points, we must declare the
location of the branch cuts. The function $\arccos(z)$ has branch
points at $z=\pm 1,\infty$, and conventionally one places the branch
cuts along the real axis from $-\infty$ to $-1$ and $+1$ to $+\infty$.
The function $\arccos(z)$ is real for real $z$ lying between $-1$
and $+1$.
For the second $\arccos$ term in (\ref{r21}), we note that the
corresponding
branch cuts lie between the points
$z_{2}$ to $z_{3}$, and $z_{1}$ to $z_{2}$.
The first $\arccos$ term has a branch cut from
$z_{2}$ to $z_{3}$.
Furthermore, due to the presence of the square root branch points
at $z_{1},z_{2},z_{3}$, we may declare a suitable branch cut for
the total action as one which extends
from $z_{1}$ to $-\infty$ along the real axis.
We note the similarity between this case and that studied in
\cite{Hartle2} and \cite{Bir1}. The only difference in the $d$-dimensional
case under study is that the location of the three branch points
depends on the dimension.
With this convention, we note that for real $z > z_{3}$, we have
real valued
Euclidean signature action, with real
volumes and real deficit angles. The action
on the first sheet is then given by (\ref{r20})-(\ref{r21})
with positive signs taken for the square root factors.

It should also be noted that the action is purely imaginary
for real $z<z_{1}$. On the first sheet, we have
\bea
F(z) &=& i\frac{d N_{d-1}(M_{d-1})}{(d-2)!}\sqrt{\frac{d-1}{2^{d-2}}}
\left[- 2 \;\mbox{\rm arcsinh}\left(\frac{1}{\sqrt{2d(d-1)(z_{2} - z)}}
\right)\right]\non\\
&+& \left. \frac{2N_{d-3}(M_{d-1})}{(d-2)!} \sqrt{\frac{d-2}{2^{d-3}}}
\sqrt{z_{1} - z}\right[2 \pi \non\\
&-&\left. \frac{d(d-1)N_{d-1}(M_{d-1})}{2N_{d-3}(M_{d-1})}
\arccos\left(\frac{z - \frac{1}{2}}{(d-1)(z-z_{2})}\right) \right],\non\\
G(z) &=& i\frac{6N_{d-1}(M_{d-1})}{d!}
\sqrt{\frac{d}{2^{d-1}}}\sqrt{z_{3} - z},
\label{r23}
\eea
where the identity,
\bea
\pi - 2\arccos(iz) = 2\arcsin(iz) = 2i\; \mbox{\rm arcsinh}(z),
\label{r24}
\eea
has been used. It is at this point that we notice the relevance of the
factor
of $2$ in the formula for the deficit angle of
a boundary $(d-2)$-simplex. Its presence allows use of the formula
(\ref{r24}), thereby rendering the action purely imaginary in the region
real $z < z_{1}$.
Thus, when the action is continued once around all three branch points,
we reach a second sheet, and the value of the action is the negative
of its value on the first sheet, as can be seen
by using the identity $\arccos(-z) = \pi - \arccos(z)$.
A continuation twice around all three
branch points returns the action to its original value.

The asymptotic behaviour of the action on its various sheets is important
for the determination of convergent contours of integration.
For large $\mid z\mid$ on the first sheet, we have
\bea
S(z,R) \sim \frac{6N_{d-1}(M_{d-1})}{d!}\sqrt{\frac{d}{2^{d-1}}}
\frac{R^{\frac{d}{2}-1}}{H^{d-2}}(R-R_{crit})\sqrt{z},
\label{r25}
\eea
where
\bea
R_{crit} = \frac{2d(d-1)N_{d-3}(M_{d-1})}{3N_{d-1}(M_{d-1})}
\sqrt{\frac{d-2}{d}}\left[2 \pi - \frac{d(d-1)N_{d-1}(M_{d-1})}{2N_{d-3}
(M_{d-1})}\arccos\left(\frac{1}{d-1}\right)\right].
\label{r26}
\eea
Thus, we see that the asymptotic behaviour of the action depends
crucially on whether the boundary edge length $R$ is greater than or less
that the critical value $R_{crit}$. It is also important to notice
that $R_{crit}$ depends only on the boundary data $N_{d-1}(M_{d-1})$
and $N_{d-3}(M_{d-1})$.

Turning now to a description of the classical extrema of the action,
we have the Regge equation of motion given by
\bea
\frac{d}{dz}S(z,R) = 0.
\label{r27}
\eea
This equation is to be solved for the value of $z$ subject to fixed
boundary data $R$. The internal lengths are then fixed by the
relation
$s_{i} = \frac{zRl^{2}}{H^{2}}$.
We also impose the physical restriction
that the boundary data $R$ should be real valued and positive.
It is straightforward to show that the Regge equation can be written as
\bea
R = \frac{F^{\prime}(z)}{G^{\prime}(z)},
\label{r28}
\eea
where the prime denotes a derivative with respect to $z$. It is thus
clear that classical solutions can exist when both $F^{\prime}$
and $G^{\prime}$ are purely real (which is the case for real
$z > z_{3}$), or when both $F^{\prime}$ and $G^{\prime}$
are purely imaginary (which is the case for real $z< z_{1}$).
Using (\ref{r26}), we then determine that for every
$0<R<R_{crit}$, there  is a real solution
of Euclidean signature at every real $z> z_{3}$. Additionally,
for every positive $R>R_{crit}$, there is a real solution of Lorentzian
signature for every real $z<z_{1}$.
These solutions occur in pairs, and by encircling all three branch
points once, one obtains a second solution with opposite value
for the action.

The main aim is to compute the wave function of the model.
In general, this problem can be tackled by first determining
the classical extrema of the action, and then searching for steepest
descent contours of integration which yield a convergent result.
In this approach, the contour depends on the
boundary value $R$.
In \cite{Hartle2}, this procedure was followed for
a four-dimensional model with a universe of $S^{3}$ topology.
However, it was also observed that the model possessed a closed
contour of integration which encircled all three branch
points, and led to a convergent wave function. Furthermore,
this closed contour of integration could be deformed to the steepest
descent contours for all values of the boundary edge length.
Thus, the
closed contour provides a contour specification which is
independent of the argument
of the wave function, namely the boundary edge length, and
as such
it constitutes a contour prescription for the model \cite{HallLou1}.

In \cite{Bir1,Bir2}, this result was extended to non-trivial topology in
four dimensions. Specifically, the case where the four-dimensional
spacetime had the structure of a cone over a boundary
of lens space topology was studied.
It was shown that the Regge action once again had three
square root branch points, and thus the analysis
of \cite{Hartle2} could be applied to the more general case.
It was observed that the key features in this application were
the simplicial minisuperspace consisting of a single
internal edge and a single boundary edge, and the cone structure
of the spacetime.

In the above, we have computed the Regge action for a cone over
a boundary of arbitrary topology in $d$ dimensions. It is clear
that, due to the analytic and asymptotic structure
of the action, the analysis
of \cite{Hartle2} also applies in this general setting.
Indeed, we have seen that the action has three square root
branch points. Furthermore, the fact that the action
is purely real for real $z > z_{3}$ and purely imaginary for
real $z < z_{1}$ ensures that the extrema of the action
are also obtained in an analogous fashion. The essential
point to be noted in the general case is
that the branch points lie along the real $z$-axis at locations which
depend on the particular dimension under study. In addition,
the critical value of the boundary edge length is determined
in terms of the number $N_{d-1}(M_{d-1})$ of $(d-1)$-simplices,
and number $N_{d-3}(M_{d-1})$ of $(d-3)$-simplices of the boundary.
The simplicity of this result allows to to quickly
survey a large number of non-trivial topologies
and determine the structure of the corresponding wave function.

\section{Three Dimensions}
Our aim in this and the following sections is to determine
the behaviour of the wave function for universes of non-trivial
topology in various dimensions. As we have seen, the nature of
the wave function is dependent on the value of the critical
length $R_{crit}$.
Here, we shall compute the value of $R_{crit}$ for
a wide variety of models. For many of the models studied, we find
that $R_{crit}$ is negative, and thus the wave
function oscillates for all positive values of the boundary edge
length. In particular, we then find
that the resulting wave function
supports a classically allowed regime, whereby the semiclassical
approximation is dominated by real Lorentzian signature
geometries.

Before dealing specifically with the three-dimensional case,
we make some general remarks.
For the models under consideration, we are taking the boundary
complex to be a simplicial manifold.
Given an $n$-dimensional simplicial complex $K$, we recall
that the star of a simplex $\sigma$ in $K$ is the collection of
simplices which contain $\sigma$, together with all their
subsimplices. The link of $\sigma$ is then the set of simplices in the
star of $\sigma$ which do not contain $\sigma$.
The simplicial complex $K$ is said to be a simplicial manifold
if and only if the link of every $k$-simplex is combinatorially equivalent
to an $(n-k-1)$-sphere \cite{RS}.

For a simplicial manifold, the following Dehn-Sommerville
relations are satisfied \cite{Kuh1}
\bea
\chi(M) &=& \sum_{i=0}^{d} (-1)^{i} N_{i},\nonumber\\
0 &=&\sum_{i=2k-1}^{d}(-1)^{i}\left(\begin{array}{c}i+1\\
2k-1\end{array}\right)N_{i},\;\;\;\;\;\mbox{\rm if}\;\; d\;\;
\mbox{\rm even},
\;\; 1\leq
k\leq\frac{d}{2},\nonumber\\
0 &=& \sum_{i=2k}^{d}(-1)^{i}\left(\begin{array}{c}i+1\\
2k\end{array}\right)N_{i},\;\;\;\;\;\mbox{\rm if}\;\; d\;\; \mbox{\rm odd},
\;\; 1\leq
k\leq\frac{d-1}{2},
\label{r29}
\eea
where $\chi(M)$ is the Euler characteristic of $M$. We shall
use these relations in the following.

In three dimensions, the most general possibility for the
topology of the boundary universe
is a genus $g$ Riemann surface $\Sigma_{g}$.
In this case, the critical length is given by
\bea
R_{crit} = \frac{4\pi}{\sqrt{3}}\frac{1}{N_{2}(\Sigma_{g})}
\left[2N_{0}(\Sigma_{g})  - N_{2}(\Sigma_{g})\right].
\label{r30}
\eea
However, since $\Sigma_{g}$ is a simplicial manifold, the Dehn-Sommerville
relations state that
\bea
N_{0}(\Sigma_{g}) - N_{1}(\Sigma_{g}) + N_{2}(\Sigma_{g}) &=&
 \chi(\Sigma_{g})
= (2 - 2g),\non\\
2N_{1}(\Sigma_{g}) - 3 N_{2}(\Sigma_{g}) &=& 0.
\label{r31}
\eea
Hence, the critical value is given by
\bea
R_{crit} = \frac{8\pi}{\sqrt{3}}\frac{1}{N_{2}(\Sigma_{g})}
(2-2g).
\label{r32}
\eea
Thus, we have the result
\bea
S^{2},\;\;\;\;\;& &R_{crit} > 0,\non\\
T^{2},\;\;\;\;\;& &R_{crit} = 0,\non\\
\Sigma_{g},\;g>1,\;\;\;\;\;& &R_{crit} < 0.
\label{r33}
\eea

\section{Four Dimensions}
Turning now to four dimensions, we begin with some specific topologies
before considering some general classes.
In \cite{Kuh1}, triangulations of $S^{2}\times S^{1}$ and $T^{3}$
have been constructed. As we have seen, the only information we need
in order to determine the critical length is the number of
simplices.
We have \cite{Kuh1}
\bea
N_{0}(S^{2}\times S^{1}) &=& 10,\non\\
N_{1}(S^{2}\times S^{1}) &=& 40,\non\\
N_{2}(S^{2}\times S^{1}) &=& 60,\non\\
N_{3}(S^{2}\times S^{1}) &=& 30,
\label{r34}
\eea
and
\bea
N_{0}(T^{3}) &=& 15,\non\\
N_{1}(T^{3}) &=& 105,\non\\
N_{2}(T^{3}) &=& 180,\non\\
N_{3}(T^{3}) &=& 90.
\label{r35}
\eea
From (\ref{r26}), we then find that
\bea
S^{2}\times S^{1},\;\;\;\;\;& & R_{crit} > 0,\non\\
T^{3},\;\;\;\;\; & & R_{crit} < 0.
\label{r36}
\eea

\subsection{Tower Construction of $\Sigma_{g}\times S^{1}$}
We are interested here in the construction of manifolds
which have a product structure $\Sigma_{g} \times S^{1}$.
Let us suppose we have a simplicial complex $K(\Sigma_{g})$.
To construct a simplicial manifold of the form $\Sigma_{g}\times S^{1}$,
we can proceed as follows. For each $2$-simplex of $K(\Sigma_{g})$,
we form a tower of $3$-simplices containing nine elements.
For example, let $[1,2,3]$ be a $2$-simplex of $K(\Sigma_{g})$
which has positive orientation.
Then the associated tower is given by
\bea
&+&[1,2,3,\tilde{1}]\nonumber\\
&+& [2,3,\tilde{1},\tilde{2}] \nonumber\\
&+& [3,\tilde{1},\tilde{2},\tilde{3}] \nonumber\\
&+& [\tilde{1},\tilde{2},\tilde{3},\hat{1}] \nonumber\\
&+& [\tilde{2},\tilde{3},\hat{1},\hat{2}] \nonumber\\
&+& [\tilde{3},\hat{1},\hat{2},\hat{3}] \nonumber\\
&+& [\hat{1},\hat{2},\hat{3},1]         \nonumber\\
&+& [\hat{2},\hat{3},1,2]               \nonumber\\
&+& [\hat{3},1,2,3],
\label{r37}
\eea
where we have introduced six new vertices $\tilde{1},\tilde{2},\tilde{3}$
and $\hat{1},\hat{2},\hat{3}$, and the relative orientations
are as shown. Thus, the number of $0$-simplices and $3$-simplices
of the simplicial manifold $K(\Sigma_{g} \times S^{1})$ are
\bea
N_{0}(\Sigma_{g}\times S^{1}) &=& 3N_{0}(\Sigma_{g}),\nonumber\\
N_{3}(\Sigma_{g}\times S^{1}) &=& 9N_{2}(\Sigma_{g}).
\label{r38}
\eea
However, the three-dimensional simplicial complex constructed
in this way is a simplicial manifold, and so we have the
Dehn-Sommerville relations
\bea
N_{0}(\Sigma_{g}\times S^{1}) - N_{1}(\Sigma_{g}\times S^{1})
+N_{2}(\Sigma_{g}\times S^{1}) - N_{3}(\Sigma_{g}\times S^{1}) &=& 0,
\nonumber\\
N_{2}(\Sigma_{g} \times S^{1}) - 2 N_{3}(\Sigma_{g}\times S^{1}) &=&0.
\label{r39}
\eea
Using (\ref{r38}), these relations provide the following information
\bea
N_{2}(\Sigma_{g}\times S^{1}) &=& 18 N_{2}(\Sigma_{g}),\nonumber\\
N_{1}(\Sigma_{g}\times S^{1}) &=& 9N_{2}(\Sigma_{g}) +
3N_{0}(\Sigma_{g}).
\label{r40}
\eea
Thus, we have determined the number of simplices $N_{i}(\Sigma_{g}
\times S^{1})$ for $i=0,1,2,3$ in terms of data associated with
$\Sigma_{g}$. This can be further simplified by recalling
that $\Sigma_{g}$ is itself a simplicial manifold, and
hence satisfies the Dehn-Sommerville relations (\ref{r31}). We
can therefore express $N_{i}(\Sigma_{g}\times S^{1})$ in terms
of the number of vertices of $\Sigma_{g}$ as follows
\bea
N_{0}(\Sigma_{g}\times S^{1}) &=& 3 N_{0}(\Sigma_{g}),\nonumber\\
N_{1}(\Sigma_{g}\times S^{1}) &=& 21N_{0}(\Sigma_{g}) -18(2 - 2g),\non\\
N_{2}(\Sigma_{g} \times S^{1}) &=& 36N_{0}(\Sigma_{g}) - 36(2 - 2g),
\nonumber\\
N_{3}(\Sigma_{g} \times S^{1}) &=& 18N_{0}(\Sigma_{g}) - 18(2 - 2g).
\label{r41}
\eea
It has been shown that the number of vertices
of a triangulation of a genus $g$ Riemann surface
satisfies the inequality \cite{Kuh1}
\bea
N_{0}(\Sigma_{g}) \geq \frac{1}{2}\left(7 + \sqrt{49 - 24(2-2g)}\right).
\label{r42}
\eea
It is then straightforward to evaluate the critical length,
yielding the result
\bea
S^{2} \times S^{1},\;\;\;\;\;& &R_{crit} > 0,\non\\
\Sigma_{g}\times S^{1},\;g\geq1\;\;\;\;\;& &R_{crit} <0.
\label{r43}
\eea

\subsection{Construction of a Connected Sum Manifold}
A second general class of topologies which can be studied
is provided by the connected sum structure.
Let us consider two simplicial manifolds $M_{1}$ and $M_{2}$.
We can construct a simplicial manifold called the connected
sum of $M_{1}$ and $M_{2}$ and denoted
$M_{1}\#M_{2}$ as follows.
We remove a single $3$-simplex, say $[0,1,2,3]$, from $M_{1}$.
We recall that a $3$-simplex is a $3$-ball with boundary $S^{2}$,
i.e.,
\bea
\partial [0,1,2,3] = [1,2,3] - [0,2,3] + [0,1,3] - [0,1,2] = S^{2}.
\label{r44}
\eea
Thus, upon removal of a single $3$-simplex, we obtain a simplicial
complex which we will denote by $M_{1}^{\prime}$, and with
$S^{2}$ boundary $\partial [0,1,2,3]$.
We also remove a single $3$-simplex, say $[\tilde{0},\tilde{1},\tilde{2},
\tilde{3}]$, from $M_{2}$, which results in a simplicial complex
denoted by $M_{2}^{\prime}$, and
with $S^{2}$ boundary $\partial[\tilde{0},\tilde{1},\tilde{2},\tilde{3}]$.
To obtain a simplicial complex for $M_{1}\#M_{2}$, we must now
identify the two $S^{2}$'s. However, in order to ensure that
the resulting structure is a simplicial manifold, this identification
is performed with the aid of a tubular neighbourhood.
A tubular neighbourhood is a simplicial complex which has
the topology of a cylinder $S^{2} \times I$, where $I$ is
the unit interval. To construct such an object, we simply
take a simplicial complex for $S^{2}$, and use the tower
construction described above, except that now we need
only three layers in the tower, viz.
\bea
K(S^{2}\times I) &=& [1,2,3,\tilde{1}] - [0,2,3,\tilde{0}]
+[0,1,3,\tilde{0}] - [0,1,2,\tilde{0}]\nonumber\\
&+& [2,3,\tilde{1},\tilde{2}] - [2,3,\tilde{0},\tilde{2}]
+[1,3,\tilde{0},\tilde{1}] - [1,2,\tilde{0},\tilde{1}]\nonumber\\
&+& [3,\tilde{1},\tilde{2},\tilde{3}] - [3,\tilde{0},\tilde{2},\tilde{3}]
+[3,\tilde{0},\tilde{1},\tilde{3}] - [2,\tilde{0},\tilde{1},\tilde{2}].
\label{r45}
\eea
It is simple to check that
\bea
\partial K(S^{2}\times I) = - \partial[0,1,2,3] + \partial [\tilde{0},
\tilde{1},\tilde{2},\tilde{3}].
\label{r46}
\eea
Thus, $K(S^{2}\times I)$ is a simplicial complex with boundary given
by the disjoint union of two $S^{2}$'s.
To obtain the simplicial manifold $M_{1}\#M_{2}$, we now
identify the $S^{2}$ boundaries of $M_{1}^{\prime}$ and
$M_{2}^{\prime}$ with the boundaries of $S^{2} \times I$.

Since our goal is to compute the critical length, we must
determine $N_{1}(M_{1}\# M_{2})$ and $N_{3}(M_{1}\# M_{2})$.
To this end, we first note that from the explicit construction
(\ref{r45}) we have
\bea
N_{0}(S^{2}\times I) &=& 8,\non\\
N_{1}(S^{2}\times I) &=& 22,\non\\
N_{2}(S^{2}\times I) &=& 28,\non\\
N_{3}(S^{2}\times I) &=& 12.
\label{r47}
\eea
This leads to the following result
\bea
N_{0}(M_{1}\# M_{2}) &=& N_{0}(M_{1}) + N_{0}(M_{2}),\non\\
N_{1}(M_{1}\# M_{2}) &=& N_{1}(M_{1}) + N_{1}(M_{2}) + 10,\non\\
N_{2}(M_{1}\# M_{2}) &=& 2N_{3}(M_{1}) + 2N_{3}(M_{2}) + 20,\non\\
N_{3}(M_{1}\# M_{2}) &=& N_{3}(M_{1}) + N_{3}(M_{2}) + 10.
\label{r48}
\eea
The number of $3$-simplices of the connected sum is easily seen
to be given by (\ref{r48}) by recalling that we have removed two
$3$-simplices in the construction of $M_{1}^{\prime}$ and
$M_{2}^{\prime}$, and added twelve through the tubular
neighbourhood.
Similarly, one notes that the tubular neighbourhood involves
the addition of ten extra $1$-simplices not already present in
either $M_{1}^{\prime}$ or $M_{2}^{\prime}$.
With this information at hand, we can now proceed and check
the value of $R_{crit}$ for several cases.
Using the tower construction of the previous section, we
obtain the following result
\bea
(S^{2}\times S^{1})\#(S^{2}\times S^{1}),\;\;\;\;\; & &R_{crit} > 0,\non\\
(S^{2}\times S^{1})\# T^{3},\;\;\;\;\;& & R_{crit} > 0,\non\\
(S^{2}\times S^{1})\# (\Sigma_{g} \times S^{1}),\;\;\;\;\; & &
R_{crit} < 0,\;\;\mbox{\rm for}\;\; g\geq 2,\non\\
(\Sigma_{g_{1}}\times S^{1})\# (\Sigma_{g_{2}}\times S^{1}),\;\;\;\;\;
& &R_{crit} < 0,\;\;\mbox{\rm for}\;\; g_{1}\geq 1,\;\; g_{2}\geq 1.
\label{r49}
\eea

In \cite{Bir1}, the wave function for a lens space boundary
was considered. A simplicial complex for a lens space of
the type $L(k,1)$ with $k\geq 2$
has been constructed in \cite{Brehm}, with the data
\bea
N_{0}(L(k,1)) &=& 2k + 7,\non\\
N_{1}(L(k,1)) &=& 2k^{2} + 12k + 19,\non\\
N_{2}(L(k,1)) &=& 4k^{2} + 20 k + 24,\non\\
N_{3}(L(k,1)) &=& 2k^{2} + 10 k + 12.
\label{r50}
\eea
We then find the critical lengths
\bea
L(k_{1},1) \# L(k_{2},1),\;\;\;\;\;& &R_{crit}>0,\;\;\;\mbox{\rm for}\;\;
(k_{1},k_{2}) = (2,2), (2,3), (2,4), (3,3), (3,4),\non\\
L(k_{1},1) \# L(k_{2},1),\;\;\;\;\;& &R_{crit}<0,\;\;\mbox{\rm otherwise}.
\label{r51}
\eea
Finally, using the tower construction of the product manifold, we
find
\bea
L(k,1) \# (S^{2}\times S^{1}),\;\;\;\;\;& &R_{crit} >0,\;\;
\mbox{\rm for}\;\; k=2,3,4,5, \non\\
L(k,1)\#(S^{2}\times S^{1}),\;\;\;\;\;& & R_{crit} < 0,\;\;
\mbox{\rm for}\;\;k\geq 6,\non\\
L(k,1)\#T^{3},\;\;\;\;\;& &R_{crit}>0,\;\;\mbox{\rm for}\;\;k=2,\non\\
L(k,1)\#T^{3},\;\;\;\;\;& &R_{crit} <0,\;\;\mbox{\rm for}\;\;
k\geq 3,\non\\
L(k,1)\# (\Sigma_{g}\times S^{1}),\;\;\;\;\;& &R_{crit} <0,\;\;
\mbox{\rm for}\;\;k\geq 2\;\;\mbox{\rm and}\;\;g \geq 2.
\label{r52}
\eea

\section{Five Dimensions}
A general construction of a simplicial complex for $S^{d-1}\times S^{1}$
and $T^{d}$ has been presented in \cite{Kuh1}.
For the case of $d=4$, we have
\bea
N_{0}(S^{3} \times S^{1}) &=& 11,\non\\
N_{1}(S^{3} \times S^{1}) &=& 55,\non\\
N_{2}(S^{3} \times S^{1}) &=& 110,\non\\
N_{3}(S^{3} \times S^{1}) &=& 110,\non\\
N_{4}(S^{3} \times S^{1}) &=& 44,
\label{r53}
\eea
and
\bea
N_{0}(T^{4}) &=& 31,\non\\
N_{1}(T^{4}) &=& 465,\non\\
N_{2}(T^{4}) &=& 1550,\non\\
N_{3}(T^{4}) &=& 1860,\non\\
N_{4}(T^{4}) &=& 744.
\label{r54}
\eea
This leads to the result
\bea
S^{3}\times S^{1},\;\;\;\;\;& & R_{crit} > 0,\non\\
T^{4},\;\;\;\;\;& & R_{crit} <0.
\label{r55}
\eea
Finally, it is interesting to study the behaviour of the critical length for
a series of triangulations of $CP^{2}$
constructed in
\cite{Kuh2}. For each $p\geq 2$, there
is a triangulation of $CP^{2}$ with the following data
\bea
N_{0}(CP^{2}_{p}) &=& p^{2} + p + 4,\non\\
N_{1}(CP^{2}_{p}) &=& 3p(p^{2} + p + 1),\non\\
N_{2}(CP^{2}_{p}) &=& 2(6p-5)(p^{2} + p + 1),\non\\
N_{3}(CP^{2}_{p}) &=& 15(p-1)(p^{2} + p + 1),\non\\
N_{4}(CP^{2}_{p}) &=&6(p-1)(p^{2} + p + 1).
\label{r56}
\eea
We find that $R_{crit} > 0$ for $p=2,3,4$, and $R_{crit} < 0$
for $p\geq 5$.
A triangulation of $CP^{2}$ with nine vertices has also been
constructed \cite{Kuh3}, such that
\bea
N_{0}(CP^{2}_{9}) &=& 9,\non\\
N_{1}(CP^{2}_{9}) &=& 36,\non\\
N_{2}(CP^{2}_{9}) &=& 84,\non\\
N_{3}(CP^{2}_{9}) &=& 90,\non\\
N_{4}(CP^{2}_{9}) &=& 36.
\label{r57}
\eea
The result is that $R_{crit}(CP^{2}_{9}) > 0$.

\section{$d$ Dimensions}
While we have discussed some examples in specific dimensions in the
preceding sections, we can now deal with a general example.
The original model discussed in \cite{Hartle2} dealt with the case of a
universe with $S^{3}$ topology. It was found that $R_{crit}$
was positive. Furthermore, for $R < R_{crit}$
the wave function was exponential in form and dominated
semiclassically by real geometry of Euclidean signature.
For $R > R_{crit}$, the wave function
was oscillatory and dominated semiclassically by real geometry
of Lorentzian signature.
It is quite straightforward to extend this analysis
to arbitrary dimensions.
A simplicial complex for $S^{d-1}$ is easily obtained as the
boundary of a $d$-simplex, namely
\bea
K(S^{d-1}) = \partial [0,1,2,...,d].
\label{r58}
\eea
A simplicial manifold is said to be $k$-neighbourly if
the following condition is satisfied \cite{Kuh1}
\bea
N_{k-1}(M) = \left( \begin{array}{c}N_{0}(M)\\k\end{array}\right).
\label{r59}
\eea
Using (\ref{r58}), one then sees
that $S^{d-1}$ is $k$-neighbourly for all $k$.
Thus, we find that
\bea
N_{d-1}(S^{d-1}) &=& (d+1),\non\\
N_{d-3}(S^{d-1}) &=& \frac{(d+1)d(d-1)}{3!}.
\label{r60}
\eea
As a result, we determine that $R_{crit} > 0$ for all $d$.
The behaviour of the wave function is also such that for
$R < R_{crit}$ and $R > R_{crit}$,
it is dominated semiclassically by real geometry
of Euclidean and Lorentzian signature, respectively.

\section{Concluding Remarks}
We have shown that the analytic structure of the
Regge action on a cone in $d$ dimensions
in simplicial minisuperspace can be obtained explicitly.
This structure allows to us determine the form of
the wave function, and we have shown that the
closed contour of integration found in four dimensions
\cite{Hartle2} is equally valid in this more general setting.
The  wave function depends crucially on the critical value
for the boundary edge length. For values of $R < R_{crit}$,
the wave function is exponential in form, and is dominated
in the semiclassical regime by real simplicial geometries
of Euclidean signature. For values $R > R_{crit}$,
the wave function has an oscillating form, and is dominated
semiclassically by
real simplicial geometries of Lorentzian signature.
We should also note that contours of integration in continuum
minisuperspace models have been studied in \cite{HallLou1} and
\cite{HallLou2}-\cite{HallMy}.
A calculation in three-dimensional Regge calculus
with torus topology was presented in \cite{LouTuck}.
The original model studied in \cite{Hartle2} was generalized
in \cite{Furihata} to
include anisotropy of the bounding universe.

Finally, we note that the behaviour of the wave function under
subdivision of the bounding universe in four dimensions was
studied in \cite{Bir1}. Such an analysis can also be performed
in $d$ dimensions. We simply have to appeal to
the general $(k,l)$ moves \cite{Pachner}, and determine their effect
on the value of $R_{crit}$.

\noindent{\bf \large Acknowledgements}\\
This work is an expanded version of an essay which received
an honourable mention
in the 1997 Gravity Research Foundation Essay Competition.
The work was supported by Forbairt grant number SC/96/603.

\end{document}